\documentclass[aps,prl,reprint,superscriptaddress]{revtex4-1}

\usepackage{graphicx}
\usepackage{bm}
\usepackage{braket}
\usepackage{amsmath}
\usepackage{longtable}
\usepackage[british]{babel}
\usepackage{verbatim} 

\begin{document}

\title{Three-Dimensional Interface Roughness in Layered Semiconductor Structures and Its Effects on Intersubband Transitions}

\author{Alex Y.\ Song}
\email[]{alexys@stanford.edu}
\affiliation{Department of Electrical Engineering, Princeton University, Princeton, NJ 08540}

\author{Rajaram Bhat}
\affiliation{Corning Incorporated, Corning, NY 14831}

\author{Pierre Bouzi}
\affiliation{Department of Electrical Engineering, Princeton University, Princeton, NJ 08540}

\author{Chung-En Zah}
\affiliation{Corning Incorporated, Corning, NY 14831}

\author{Claire F.\ Gmachl}
\affiliation{Department of Electrical Engineering, Princeton University, Princeton, NJ 08540}

\date{\today}

\begin{abstract}

A general model for treating the effects of three dimensional interface roughness (IFR) in layered semiconductor structures has been derived and experimentally verified. Configurational averaging of the IFR potential produces an effective grading potential in the out-of-plane direction, greatly altering the energy spectrum of the structures.
 IFR scattering self-energy is also derived for the general case; when IFR is strong, its scattering effect is shown to dominate over phonon interaction and impurity scattering. When applied to intersubband transitions, the theoretical predictions explain the experimental observation of the anomalous energy shift and unusual broadening of the ISB transitions in III-Nitride thin-layered superlattices.

\end{abstract}


\maketitle

Heterointerface is an essential structure
in a wide range of research fields
\cite{miao_PRL2012GaN,takagi_PRB2013_O-IO,konig_Science2007_HgTe,king_PRL2012_perovskite_subband,kuch_NM2006tuningMagMulti}. It is frequently associated with significant interface roughness (IFR),
such as in III-nitride heterostructures, II-VI thin films including ZnSe or HgTe, perovskite quantum wells (QWs), and magnetic multilayers
\cite{almeida_PRB2014fluctuations,boullay_PRB2011_Perovskiteroughness,Edmunds_Malis2013PRB,nakagawa_NM2006PerovskiteInterface,santamaria_PRL2002scalingIFRMagMulti}.
IFR plays an vital role in determining the transport and optical properties of such structures. 
Traditional approaches to the IFR  effects have been based on the premise of near-perfect interfaces. For example, in study of IFR scattering, a two-dimensional (\mbox{2-D}) IFR random potential is assumed which only appears on the average interface plane. This \mbox{2-D} approximation has been been universally applied in studies ranging from condensed matter heterointerfaces to the Casimir effect \cite{chevoir_PRB1993deltaz,Ando1982RevModPhys}. In semiconductor samples with very high growth quality, 
a QW layer can even be regarded as adjoining lateral regions with different thicknesses but zero IFR \cite{brunner_APL1994MLFluctu,leosson_PRB2000MLFluctu}. While these treatments are valid when IFR is sufficiently small, their general feasibility remains unverified.
In the mean while, various interesting phenomena have been observed in the study of subband structures in QWs, with examples including
anomalous intersubband (ISB) transition energy shift between experimental observations and theoretical predictions, unusual broadening in the ISB transition spectra, topological phase transitions, etc \cite{Bayram_Razeghi2009,Edmunds_Malis2013PRB,Julien_JAP2013_systematicISB,miao_PRL2012GaN}.
Understanding the subband structure and the associated ISB transitions are essential for further scientific study and devices implementations \cite{lee_PRB2009_ISB}, and it is of interest to revisit the theoretical model for treating the effects of IFR.

Here, we develop a formalism to accommodate IFR in the general scenario.
The generic stochastic form of the IFR potential with explicit \mbox{3-D} dependence is retained, i.e.\ dropping the \mbox{2-D} approximation.
Configurational average of the IFR potential produces the effective interface grading (EIG) on the lowest order, which significantly alters the energy spectrum. We also derive the IFR scattering self-energy in the general case.
The IFR scattering is shown to be dominant over longitudinal optical (LO) phonon and impurity scattering when strong IFR exists. In ISB transitions, the IFR scattering leads to extra broadening in the transition spectra. These predictions are confirmed by experimental examination of the ISB transitions in III-nitride superlattices. And the calculation also allows quantified extraction of the roughness parameters.

\begin{figure}
  \centering
  \includegraphics[width=\linewidth]{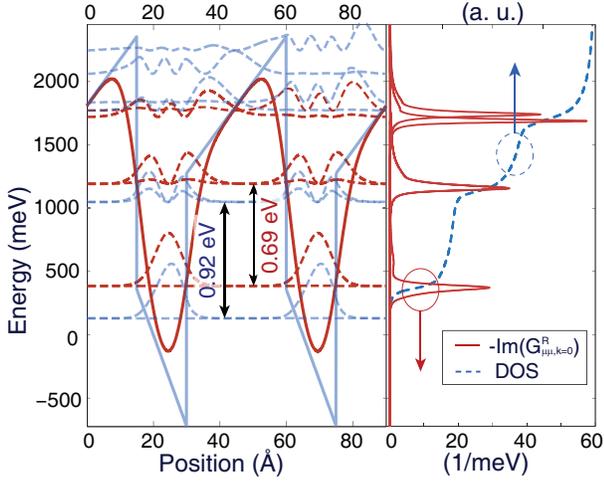}\\
  \caption{Subband structure of a 100-period GaN 1.5 nm / AlN 3 nm superlattice, with Si doping of 1.6 $\times$ $10^{19}$ cm$^{-3}$. An Al$_{0.67}$Ga$_{0.33}$N bottom template is employed.
  Left: an example of the original superlattice potential (blue) and that containing $\Sigma^g$ (red), assuming $\eta=5.5\,\textrm{\AA}$ and $\lambda=7\,\textrm{\AA}$. The corresponding WS and proper-WS states are also plotted.
  Right:
  $\Im(G^{R}_{\mu\mu,{\bm k}=0})$ and the total density of states (DOS) obtained in the full calculation. Both the wave functions and the energy spacings of the proper-WS states are altered compared to the original.
  }
  \label{Graded}
\end{figure}

We perform a full quantum modelling of the effects of \mbox{3-D} IFR within the framework of non-equilibrium Green's functions
 \cite{Wacker1998,Wacker2002PhysRevB.66.245314,RammerRMP1991quantum}. The model is explained as follows.
The general Hamiltonian can be written as $ H = H_0 + H_{i} + H_{ifr}$. The non-interacting $H_{0}$ includes the effective mass Hamiltonian within k$\cdot$p theory \cite{PhysRevB.48.8102}, i.e.\ the superlattice potential assuming ideally smooth interfaces. The nonlinear spontaneous and piezoelectric polarization potentials are also contained in $H_0$.
$H_0$ is separable 
and can be diagonized straightforwardly, whose eigen-system is known as the Wannier-Stark (WS) basis.
The wave functions of a WS state is expressed as
$1/\sqrt{A} e^{i{\bm k}{\bm r}} \psi_{\mu}(z)$, where the index $\mu$ represent the confined states in the out-of-plane direction $z$, ${\bm r}$ stands for the in-plane coordinates, and ${\bm k}$ represents the in-plane momentum. An example of such a structure is shown in Fig.~\ref{Graded} (left part, blue curves). The interacting term $H_{i}$ 
 includes electron-phonon interaction, impurity scattering and electron-electron interaction, respectively. $H_{ifr}$ represents the IFR random potential. The matrix element of $H_{ifr}$ in the WS basis is denoted as $V^{ifr}_{\mu\nu}({\bm k},{\bm k}')$.

The characteristics of the structure are obtained by solving the Dyson equation
\begin{equation}\label{Gc}
\begin{aligned}
   (\epsilon&-h^0_{\mu\nu,{\bm k}}-h^{MF}_{\mu\nu,{\bm k}})\,G_{\nu{\bm k},\mu'{\bm k}'\!}(\epsilon) \\
   = &\:\delta_{\mu,\mu'}\delta_{{\bm k},{\bm k}} \!+( \Sigma^{e-ph}_{\mu{\bm k},\nu{\bm k''}\!}(\epsilon) +\Sigma^{imp}_{\mu{\bm k},\nu{\bm k''}\!}(\epsilon))\,G_{\nu{\bm k''}\!\!,\mu'{\bm k}'\!}(\epsilon) \\
  & + (\Sigma^{g}_{\mu{\bm k},\nu{\bm k''}}\!+ \Sigma^{s}_{\mu{\bm k},\nu{\bm k''}\!}(\epsilon) ) \,G_{\nu{\bm k}''\!\!,\mu'{\bm k}'\!}(\epsilon)
\end{aligned}
\end{equation}
where $\epsilon$ is energy, $h^0$ is the matrix element of $H^0$, $\delta$ is the Kronecker delta, $G$ is the retard Green's function, $\Sigma$ represents various self-energies. Repeated indices are summed. The terms in (\ref{Gc}) are explained as follows. $\Sigma^{e-ph}$ and $\Sigma^{imp}$ are self-energies of electron-phonon interaction and impurity scattering, respectively. They are calculated with the Fock-type diagram in self-consistent Born Approximation (SCBA) \cite{RammerRMP1991quantum}.
The electron - electron Coulomb interaction is treated in the mean field approximation, i.e.\ Poisson equation; and $h^{MF}$ is the mean-field potential calculated from
\begin{equation}\label{Poisson}
 \partial^2_z h^{MF}(z) \\
 = \frac{e}{\varepsilon}(  2i\sum_{\mu,{\bm k}}\int\frac{d\epsilon}{2\pi}G^<_{\mu\mu{\bm k}}(\epsilon)\psi^2_{\mu}(z) - \rho_d(z)  )
\end{equation}
where $\varepsilon$ is the permittivity, $G^<(\epsilon)=-2i\!\cdot\!n(\epsilon)\Im G^R(\epsilon)$, $n(\epsilon)$ is the Fermi-Dirac distribution, and $\rho_d(z)$ is the density of the ionized impurities.

$\Sigma^{g}$ and $\Sigma^{s}$ in the Dyson equation (\ref{Gc}) are IFR originated self-energies:
\begin{equation}\label{SelfEirg}
\begin{aligned}
  \Sigma^{g}_{\mu{\bm k},\nu{\bm k''}} &= \langle V^{ifr}_{\mu,\nu} ({\bm k} , {\bm k}'') \rangle\\
\end{aligned}
\end{equation}
\vspace*{-12pt}
\begin{equation}\label{SelfEirs}
   \Sigma^{s}_{\mu{\bm k},\nu{\bm k''}}(\epsilon)
   =
  \langle V^{ifr}_{\mu\alpha} ({\bm k},{\bm k}_1) V^{ifr}_{\beta\nu} ({\bm k}_2,{\bm k''})\rangle
     G_{\alpha{\bm k}_1,\beta{\bm k}_2}(\epsilon)
\end{equation}
where angle brackets $\langle\rangle$ are understood as the configurational average. $\Sigma^{g}$ corresponds to the ``single leg'' diagram of random potential scattering.
$\Sigma^{s}$  is the scattering self-energy induced by IFR, which is handled within SCBA.
The expressions of $\Sigma^{g}$ and $\Sigma^{s}$ depend on $V_{ifr}$.

Generically, the IFR stochastic potential $V_{ifr}$ is a \mbox{3-D} function: 
\begin{equation}\label{Virp}
  V_{ifr}(z,{\bm r}) = \sum_{j} \delta E_j (\theta (\tilde{z}_j-\xi_j({\bm r})) - \theta (\tilde{z}_j))
\end{equation}
where $\delta E_j$ is the band offset at the $j^{\textrm{th}}$ interface, $z_j$ is the $j$'s interface position, $\tilde{z}_j\doteq z-z_j$, $\theta$ is the Heaviside function, and $\xi_j({\bm r})$ is the interface fluctuation at the in-plane location ${\bm r}$ of the $j^{\textrm{th}}$ interface. In (\ref{Virp}) we retain the original form of the IFR stochastic potential with explicit \mbox{3-D} dependence, 
and the approximation of a purely \mbox{2-D} IFR potential is dropped.

$\xi_j(\bm r)$ is a Gaussian random process \cite{Khurgin_APL2008_roughness} with a probability distribution density $f_{\xi}(\zeta)$ and a correlation as
\begin{equation}\label{xi}
  \begin{aligned}
  f_{\xi}(\zeta) = \frac{ e^{-\zeta^2/2\eta^2}}{\sqrt{2\pi}\eta},\quad
   \langle\xi_j({\bm r}_1)\,\xi_j({\bm {r}_2})\rangle  = \eta^2e^{-r^2/\lambda^2}\\
  \end{aligned}
\end{equation}
where $\eta$ is the roughness height, $\lambda$ is the correlation length, and $r=|{\bm r}_1 - {\bm r}_2|$. Furthermore, the joint probability density at $\xi_j({\bm r}_1)=\zeta$ and $\xi_j({\bm r}_2)=\zeta'$ is
\begin{equation}\label{Jointpdf}
  f^{(2)}_{\xi_j,r}(\zeta,\zeta') = \frac{1}{2\pi \sqrt{det(C)}} e^{-({\zeta,\zeta'})C^{-1}({\zeta,\zeta'})^T}
\end{equation}
where $C=\eta^2(I+e^{-r^2/\lambda^2}\sigma_x)$ is the correlation matrix,
$I$ and $\sigma_x$ are the identity matrix and the $x$-Pauli matrix, respectively. 

With the original \mbox{3-D} form of IFR potential retained in (\ref{Virp}), $\Sigma^g$ can be expressed as \footnote{See the supplementary material for the detailed derivation of $\Sigma^{g,s}$.}
\begin{equation}\label{SelfEirgeff}
\begin{aligned}
  \Sigma^{g}_{\mu{\bm k},\nu{\bm k''}}
  =V^{g}_{\mu {\bm k}, \nu{\bm k}''} - V^0_{\mu{\bm k}, \nu{\bm k}''}
\end{aligned}
\end{equation}
where 
\begin{equation}\label{Vg}
  \begin{array}{c}
     V^{g}_{\mu {\bm k}, \nu{\bm k}''}  \\
     V^0_{\mu{\bm k}, \nu{\bm k}''}
   \end{array}
  \!\! = 4\pi^2\delta_{{\bm k}, {\bm k''}} \!\!\int \!dz \sum_{j} \delta E_{j\!} \begin{array}{c}
                                                                                                            F_{\xi}(\tilde{z}_j) \\
                                                                                                            \theta (\tilde{z}_j)
                                                                                                          \end{array}\!
                                        \psi^*_{\mu} (z) \psi_{\nu} (z)
\end{equation}
and $F_{\xi}(\tilde{z}_j)=1/2(1+erf(\tilde{z}_j/\sqrt{2}\eta))$ is the cumulative probability distribution. $erf$ is the error function.

If a \mbox{2-D} IFR potential 
is assumed, the ``single leg'' diagram of $\Sigma^{g}$ would produce an universal constant zero, thus has no physical effect.
We recognize $V^0$ as precisely the unperturbed superlattice potential with ideally smooth interfaces.
$\Sigma^g$ can be merged into $h^0$, retaining the separability of the Hamiltonian.
We call the basis formed by the eigenstates of $H_0+\Sigma^g$ the ``proper-WS'' basis.

We have plotted an example of the superlattice potential added with $\Sigma^g$ in Fig.~\ref{Graded} (left part, red curves).
It is observed that inclusion of $\Sigma^g$ leads to an effective interface grading. As a result, the depths of the wells are reduced, causing a narrowing in the proper-WS subband energy spacing.

Without IFR, the polarization charges are localized on the ideal interface planes. However, due to roughness they are now distributed in $z$.
This induces a correction to the polarization fields close to the interfaces, followed by a small change of $\leq$~30~meV in the ISB transition energies in these samples. This effect by itself does not explain the large energy ISB transition energy shift, but is included for accuracy in our calculation.

The IFR scattering self-energy $\Sigma^{s}$ introduced in (\ref{SelfEirs}) plays a vital role in the broadening and transport characteristics in the superlattice, and it can also contribute to the energy renormalization of the subbands.
Based on the \mbox{3-D} form of the IFR stochastic potential, $\Sigma^{s}$ can be expressed as \cite{Note1}
\begin{equation}\label{SelfEirs1}
\begin{aligned}
   \Sigma^{s}_{\mu{\bm k}\nu{\bm k''}}(\epsilon)
  = &\int\!d^2 {\bm p} \sum_{j} \frac{\delta E_j^2}{4\pi^2}\int\!d^2 {\bm r} e^{-i {\bm p} {\bm r}} \!\iint\!d z dz'  \\
   & \cdot sgn(zz')\iint_{(\zeta,\zeta')\in D} d\zeta d\zeta' f^{(2)}_{\xi, r}(\zeta,\zeta')  \\
   & \cdot  \mathcal{F}_{\mu\alpha\beta\nu}(z,z') \cdot G_{\alpha,{\bm k}-{\bm p},\beta,{\bm k}''+{\bm p}}(\epsilon)
  \end{aligned}
\end{equation}
where $\mathcal{F}_{\mu\alpha\beta\nu}(z,z')=\psi_\mu^*(z) \psi_\alpha(z) \psi_\beta^*(z') \psi_\nu(z')$, $f^{(2)}_{\xi, r}(\zeta,\zeta')$ is the joint probability distribution found in (\ref{Jointpdf}), and the domain of integration is
\begin{equation}\label{D}
  \begin{aligned}
  &D= \{  \begin{array}{ll}
         (-\infty,\tilde{z}_j), & \tilde{z}_j<0 \\
         (\tilde{z}_j,\infty), & \tilde{z}_j>0 \\
         \end{array} \times \{ \begin{array}{ll}
         (-\infty,\tilde{z}'_j), & \tilde{z}'_j<0 \\
         (\tilde{z}'_j,\infty), & \tilde{z}'_j>0\\ \end{array}
  \end{aligned}
\end{equation}

To retrieve the energy structure of the superlattices, the Dyson equation (\ref{Gc}) and the Poisson equation (\ref{Poisson}) are calculated iteratively. An example of the calculated imaginary part of the Green's functions $\Im(G^{R}_{\mu\mu,{\bm k}=0})$  are plotted in Fig.~\ref{Graded} (right part), which represents the respective density of states (DOS) for each proper-WS subband state. Broadening in each subbands  is discernable. The total DOS, $\sum_k \!2\Im(G^{R}_{\mu\mu,{\bm k}})$ is also plotted in Fig.~\ref{Graded}. The staircase shape of the total DOS is a signature of a \mbox{2-D} system, with the onset of each step corresponding to each proper-WS subband.
Based on the calculated Green's functions, the ISB transition spectrum can be generated by calculating the real part of the conductivity using the Kubo formula \cite{RammerRMP1991quantum,Ando1978}\footnote{Here we retain full $G$ instead of the approximation of using the ``on-shell'' $G_0$ }.

\begin{figure}
  \centering
  \includegraphics[width=\linewidth]{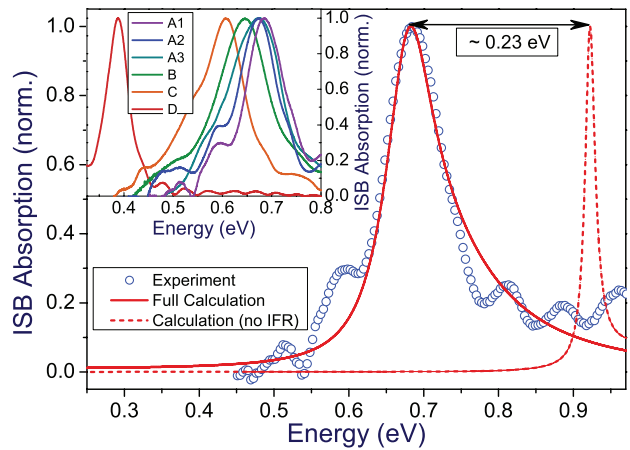}\\
  \caption{Blue circles: measured ISB absorption spectrum of design A1, obtained by taking the transverse magnetic (TM) over transverse electric (TE) absorption, and normalized. Red solid and dashed curves: calculated ISB absorption spectrum of A1 with and without the effects of \mbox{3-D} IFR, respectively. Fitted roughness parameters of $\eta = 5.6\,\textrm{\AA}$ and $ \lambda = 4.3\,\textrm{\AA}$ are used.
  Inset: normalized ISB absorption spectra of all designs at room temperature. Interference patterns
  are discernable in the spectra.
  }
  \label{Absorption}
\end{figure}

For a systematic study of the effect of \mbox{3-D} IFR in the subband structure, we have designed, grown and characterized a series of  GaN/Al(Ga)N superlattices with varying parameters listed in Table \ref{tab_samples}.
All samples are grown by metal organic chemical vapor deposition (MOCVD) on c-plane sapphire. Multi-layered templates are employed, with the final layer being strain relaxed Al$_x$Ga$_{1-x}$N matching to the average Al concentration in the active layers, ensuring balanced strain in the superlattices.
The average layer thicknesses are controlled within $\pm$ 3.5\% of the designed values.
Multiple samples ($\geq$ 3) are grown for the same designs to ensure repeatability.
To compare to the theoretical results, an experimental estimation of the roughness height is obtained through characterization of the top surface morphology.
To this end, atomic force microscope (AFM) measurements are performed at multiple locations and over multiple wafers. The result gives an average roughness height of 6~$\textrm{\AA}$ with a standard error of $\pm$~2~$\textrm{\AA}$.

\begin{table}[hb]
\caption{\label{tab_samples}%
Designed III-nitride superlattice structures. The number of periods is 100. Si doping introduced in the wells.
}
\begin{ruledtabular}
\begin{tabular}{cccc}
\textrm{Sample}&
\textrm{GaN (nm)}&
\textrm{AlN (nm)}&
\textrm{Doping ($\times10^{19}$ cm$^{-3}$)}\\
\colrule
A1 & 1.5 & 3.0 & 0.8\\
A2 & 1.5 & 3.0 & 1.6\\
A3 & 1.5 & 3.0 & 3.2\\
B & 2.0 & 5.0 & 1.6\\
C & 3.0 & 5.0 & 1.6\\
D & 3.0 & 3.0 (Al$_{0.6}$Ga$_{0.4}$N) & 0.8 \\
\end{tabular}
\end{ruledtabular}
\end{table}

\begin{figure}
  \centering
  \includegraphics[width=\linewidth]{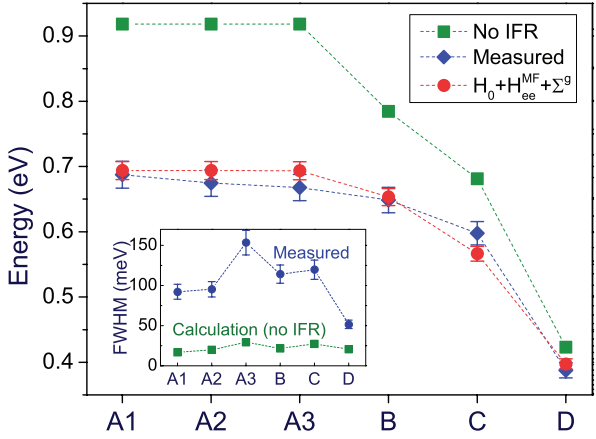}\\
  \caption{Green: calculated ISB transition energy without IFR effects for each design. Blue: Measured ISB peak transition energies. Red: calculation with $H_0 + \Sigma^g + H^{MF}$.
  The discrepancy between the the baseline calculation and the experimental results is accounted for by including the effective grading potential.
  Inset: Measured FWHM of the ISB transitions (purple) and that calculated without the effects of IFR (green). The expected broadening without considering the \mbox{3-D} IFR scattering is merely 20\% - 40\% of the measured FWHM.
  }
  \label{Wavelengths}
\end{figure}

In Fig.~\ref{Absorption} we have plotted the measured ISB absorption spectrum of design A1 (blue circles), the baseline spectrum calculated without the effects of IFR (dashed red curve) and that obtained in the full calculation (red solid curve).
The observed ISB absorption only appears in the transverse magnetic polarization, which is a signature of ISB transition.
The relevant material parameters used in the calculation are found in Ref. \cite{HM2009,*JP_Book2007nitride}. The measured peak transition energy at 0.69 eV display a $\sim$ 0.23 eV red-shift from the baseline calculation. Conversely, with the effects of \mbox{3-D} IFR included, the full calculation successfully reproduces both the peak transition energy and the broadening of the experimental result.
The the measured ISB absorption spectra of all the designs in Table \ref{tab_samples} are plotted in the inset of Fig.~\ref{Absorption}. 
The measured peak transition energies span 0.39 eV - 0.69 eV.
A summary of the measured peak energies (purple) and
that of the baseline calculation without considering the IFR effects (green) are shown in Fig.~\ref{Wavelengths}.
To demonstrate the effect of the effective grading potential, the peak ISB transition energies calculated with $H_0+\Sigma^g+H^{MF}$ and an universally fitted roughness height of 5.5~$\textrm{\AA}$ are also shown in Fig.~\ref{Wavelengths} (red).
All measured ISB transitions exhibit clear red-shift of up to $\sim$ 25\% compared to the results from the baseline calculation. The deviation in the average layer thickness of $\pm$ 3.5 \% can only lead to a energy shift of less than $\pm$ 20 meV, which can not account for such a significant discrepancy.
The electron-electron and electron-ionized impurity interactions in these structures merely contribute to $\leq$~5~meV of energy shift in the subbands, and is thus not the main reason for the observed discrepancy either.
Contrarily, the calculation equipped with $\Sigma^g$ immediately brings the predicted ISB transition energies close to the experimental results, which is a clear evidence of the effect of the \mbox{3-D} IFR. This reaffirms the reduction of energy spacings between the proper-WS subbands observed in Fig.~\ref{Graded}.

\begin{figure}
  \centering
  \includegraphics[width=\linewidth]{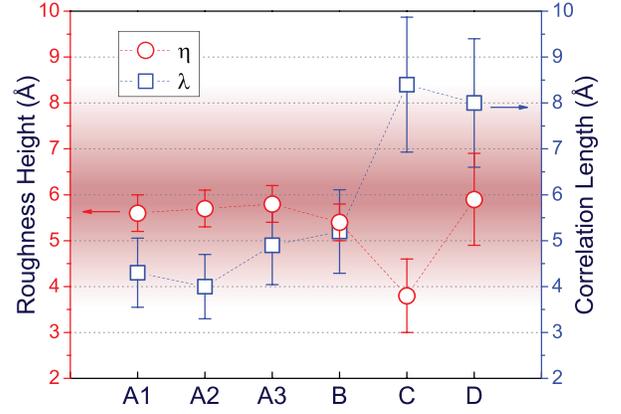}\\
  \caption{Roughness heights $\eta$ (red circles) and correlation lengths $\lambda$ (blue triangles) obtained from fitting the calculated ISB transition spectra to the experimental results.  The shaded region represent the experimental estimation of the roughness height, 6~$\pm$~2~$\textrm{\AA}$.
  }
  \label{FittedRH_CL}
\end{figure}

The measured full width at half maximum (FWHM) of the ISB transitions, originated from the broadening of the proper-WS subbands, ranges from 50 meV to 150 meV (13\% - 23\% of the transition energy) in different designs. A summary is shown in the inset of Fig.~\ref{Wavelengths}. Such values are significantly larger than those found in the III-phosphide or -arsenide material system.  For a comparison, the calculated FWHM without the IFR effects are also plotted in the inset.
The resulting broadening is merely 20\% - 40\% of the measured FWHM, clearly indicating the importance of the missing factor of the IFR scattering.
With the full theoretical model developed above, one can extract the roughness height $\eta$ and the correlation length $\lambda$ in each sample by fitting to the peak position and the FWHM of the ISB transition spectra.
A summary of the extracted $\eta$ and $\lambda$ is shown in Fig.~\ref{FittedRH_CL}.
The experimental estimation of the roughness height $\eta$ is also indicated in the shaded region in Fig.~\ref{FittedRH_CL}. All fitted $\eta$ reside within the range of experimental results.
For the correlation length $\lambda$, proper experimental measurement methods are still under discussion, with large uncertainty found in the reported values (14~$\textrm{\AA}$ $\sim$ 120~$\textrm{\AA}$) for the more studied materials \cite{Gold_APL2008Parameters,Khurgin_APL2008_roughness,Tsui_APL2007transport,Quang_APL2009correlation}.
It is worthy to note that in this work the energy shift and the broadening provide two constraints, which enables simultaneous fitting of $\eta$ and $\lambda$.
The resulting correlation length $\lambda$ ranges from 4~$\textrm{\AA}$ to 10~$\textrm{\AA}$.
The variation among samples A-D is understandable since they have different structure designs and are grown on templates with different material compositions.
The correlation lengths found here are generally smaller than those in III-phosphide or -arsenide material system;
this meets our expectation given that the interfaces in III-nitride materials are known to be considerably rougher.

\begin{figure}
  \centering
  \includegraphics[width=\linewidth]{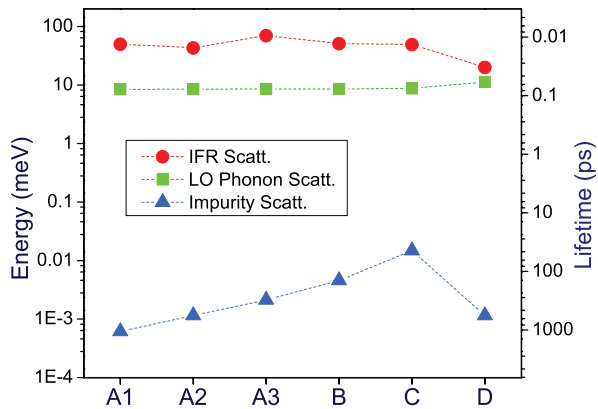}\\
  \caption{Scattering lifetimes due to \mbox{3-D} IFR (red), LO phonons (green) and impurities  (blue) for all designs. Units are given in both ps (right axis) and the corresponding energy in meV (left axis).}
  \label{lifetimes}
\end{figure}

In Fig.~\ref{lifetimes} we have plotted the semi-classical lifetimes of \mbox{3-D} IFR, LO phonon and impurity scattering to illustrate their relative importance in the proper-WS subband broadening \cite{Wacker2002PhysRevB.66.245314}.
As is shown, LO phonon scattering lifetimes are typically $\sim$ 0.05 ps, while that of impurity scattering is $>$ 50 ps. Again, clear dominance of the IFR scattering of $\sim$ 0.01 ps is observed, a result of strong \mbox{3-D} IFR in thin QWs.

In summary, we have developed a general model for treating the effect of \mbox{3-D} IFR in thin layered semiconductor structures. An effective grading potential is predicted in the model, which significantly alters the energy spectrum and leads to a narrowing in the subband spacing.  The IFR scattering self-energy is also derived for the general case, and the IFR scattering mechanism is shown to be dominant over phonon and impurity scattering.
Through the full calculation, the anomalous energy shift and the significant broadening observed in the ISB transitions are explained.
Beyond the optical transitions, the results in this work are also applicable to the vertical as well as in-plane transport studies.
Investigations of similar IFR effects in structures of other dimensions such as quantum wires and quantum dots are also enabled.
Equipped with the quantitative results of this work, the ISB light emission in III-nitride superlattices in the mid-infrared wavelength range is realized \cite{song_CLEO2014mid,song_QCE}.

\begin{acknowledgments}
This work is supported in part by MIRTHE (NSF-ERC). The authors would like to thank Dr. Joesph Maciejko for valuable discussions.
\end{acknowledgments}


%

\end{document}


\title{Supplementary Material: Derivation of Interface Roughness Self-Energies $\Sigma^{g,s}$}

\author{Alex Y.\ Song}
\email[]{alexys@stanford.edu}
\affiliation{Department of Electrical Engineering, Princeton University, Princeton, NJ 08540}

\author{Rajaram Bhat}
\affiliation{Corning Incorporated, Corning, NY 14831}

\author{Pierre Bouzi}
\affiliation{Department of Electrical Engineering, Princeton University, Princeton, NJ 08540}

\author{Chung-En Zah}
\affiliation{Corning Incorporated, Corning, NY 14831}

\author{Claire F.\ Gmachl}
\affiliation{Department of Electrical Engineering, Princeton University, Princeton, NJ 08540}

\date{\today}

\begin{abstract}

\end{abstract}


\maketitle

\section{Derivation of $\Sigma^g$}

Self-energy $\Sigma^g$ is the first order configurational average of the interface roughness (IFR) stochastic potential:
\begin{equation}\label{SelfEirg}
\begin{aligned}
  \Sigma^{g}_{\mu{\bm k},\nu{\bm k''}} &= \langle V^{ifr}_{\mu,\nu} ({\bm k} , {\bm k}'') \rangle\\
\end{aligned}
\end{equation}
where $\mu,\nu$ are indices of the Wannier-Stark (WS) basis, ${\bm k}$ is in-plane momentum. The generic form of the IFR random potential has \mbox{3-D} dependence on the interface fluctuation,
\begin{equation}\label{Virp}
  V_{ifr}(z,{\bm r}) = \sum_{j} \delta E_j (\theta (\tilde{z}_j-\xi_j({\bm r})) - \theta (\tilde{z}_j))
\end{equation}
where $z$ is the out-of-plane position, ${\bm r}$ is in-plane position, $\delta E_j$ is the band offset at the $j^{{\rm th}}$ interface, $\theta$ is the Heaviside function, $\tilde{z}_j\doteq z-z_j$, $z_j$ is the location of the $j^{{\rm th}}$ interface, $\xi_j({\bm r})$ is the interface fluctuation at ${\bm r}$.
Traditional approach invoke the \mbox{2-D} approximation for simplicity \cite{chevoir_PRB1993deltaz},
\begin{equation}\label{Virp2D}
  V_{ifr,2D}({z,\bm r}) = \sum_{j} \delta E_j \delta(\tilde{z}_j)\xi_j({\bm r})
\end{equation}
but such a \mbox{2-D} approximation is removed here.

The Gaussian properties of $\xi_j(\bm r)$ include (expression (6) and (7) in the manuscript)\cite{Khurgin_APL2008_roughness}:
\begin{equation}\label{xi}
  \begin{aligned}
  f_{\xi}(&\zeta)= \frac{ e^{-\zeta^2/2\eta^2}}{\sqrt{2\pi}\eta},\quad
   \langle\xi_j({\bm r}_1)\,\xi_j({\bm {r}_2})\rangle  = \eta^2e^{-r^2/\lambda^2}\\
  &   f^{(2)}_{\xi_j,r}(\zeta,\zeta') = \frac{1}{2\pi \sqrt{det(C)}} e^{-({\zeta,\zeta'})C^{-1}({\zeta,\zeta'})^T}
  \end{aligned}
\end{equation}
where $f_{\xi}(\zeta)$ is the probability distribution density,  $\langle\xi_j({\bm r}_1)\,\xi_j({\bm {r}_2})\rangle$ is the correlation,  $f^{(2)}_{\xi_j,r}(\zeta,\zeta')$ is the joint probability density;  $\eta$ is the roughness height, $\lambda$ is the correlation length, and $r=|{\bm r}_1 - {\bm r}_2|$. $C=\eta^2(I+e^{-r^2/\lambda^2}\sigma_x)$ is the correlation matrix, and $I$ and $\sigma_x$ are the identity matrix and the $x$-Pauli matrix, respectively.

The matrix element of the IFR potential in the WS basis  writes
\begin{equation}\label{VirpWS}
\begin{aligned}
  V^{ifr}_{\mu\nu}({\bm k},{\bm k}')  = & \int dz \sum_{j} \frac{\delta E_j}{A}\int d^2 {\bm r}  e^{-i({\bm k}-{\bm k}'){\bm r}}   \\
  &(\theta (\tilde{z}_j-\xi_j({\bm r})) - \theta (\tilde{z}_j)) \psi^*_{\mu} (z) \psi_{\nu} (z)
  \end{aligned}
\end{equation}
where A is the in-plane area of the sample, $\psi_\mu(z)$ represent the wave functions of the WS states.

As a remark, here if one assumes the wave functions are constant over the interface as an assumption, i.e.\ $\psi(z)\approx \psi(z_j)$, then the matrix element of the IFR potential in (\ref{VirpWS}) simplifies into
\begin{equation}\label{TradIRS}
  V^{ifr}_{\mu \nu} ({\bm p})  \approx
   \sum_{j}\frac{ \delta E_j }{A}\int d^2 {\bm r} e^{-i{\bm p}{\bm r}} \xi_j({\bm r}) \psi^*_{\mu} (z_j) \psi_{\nu} (z_j)
\end{equation}
which is the classical expression used to calculate interface roughness scattering \cite{Khurgin_APL2008_roughness}. The approximation of constant wave functions is equivalent to the \mbox{2-D} approximation of IFR; the expression in (\ref{TradIRS}) can be obtained identically with either assumption. Of course such approximations are only valid in case of very small IFR.

The self-energy $\Sigma^g$ is the configurational average of (\ref{VirpWS}), thus
\begin{equation}\label{SelfEirgF}
\begin{aligned}
  \Sigma^{g}_{\mu,{\bm k},\nu,{\bm k''}} =&\int dz \sum_{j} \frac{\delta E_j}{A} \int d^2 r e^{-i({\bm k}-{\bm k''}){\bm r}} \\
  &  (\langle \theta (\tilde{z}_j-\xi_j({\bm r}))\rangle - \theta (\tilde{z}_j)) \psi^*_{\mu} (z) \psi_{\nu} (z)
\end{aligned}
\end{equation}
in which
\begin{equation}\label{erf}
\begin{aligned}
  \langle\theta (\tilde{z}_j-\xi_j({\bm r}))\rangle &= \int_{-\infty}^{\infty} \theta(\tilde{z}_j-\zeta) f_{\xi} (\zeta)d\zeta\\
                                             &= \int_{-\infty}^{z-z_j} \frac{1}{2 \pi \eta} e^{-\zeta ^2 / 2\eta^2} d\zeta\\
                                             &=F_{\xi}(\tilde{z}_j)
\end{aligned}
\end{equation}
where $f_{\xi}(\zeta)$ is the probability density of $\xi_j(r)$, $F_{\xi}(\tilde{z}_j)=1/2(1+erf(\tilde{z}_j/\sqrt{2}\eta))$ is the cumulative probability distribution, $erf$ is the error function, $\eta$ is the roughness height. We notice that expression (\ref{erf}) is independent of ${\bm r}$. Thus the ${\bm r}$ integration in (\ref{SelfEirgF}) produces a delta function in momentum. Plugging (\ref{erf}) into (\ref{SelfEirgF}), we obtain the final expression of $\Sigma^g$,
\begin{equation}\label{SelfEirgeff}
\begin{aligned}
  \Sigma^{g}_{\mu{\bm k},\nu{\bm k''}}
   =V^{g}_{\mu {\bm k}, \nu{\bm k}''} - V^0_{\mu{\bm k}, \nu{\bm k}''}
\end{aligned}
\end{equation}
where 
\begin{equation}\label{Vg}
  \begin{array}{c}
     V^{g}_{\mu {\bm k}, \nu{\bm k}''}  \\
     V^0_{\mu{\bm k}, \nu{\bm k}''}
   \end{array}\!\!\!
   = 4\pi^2\delta_{{\bm k}, {\bm k''}} \!\!\int \! dz \!\sum_{j} \delta E_j \!\!\begin{array}{c}
                                                                                                            F_{\xi}(\tilde{z}_j) \\
                                                                                                            \theta (\tilde{z}_j)
                                                                                                          \end{array}\!
                                        \psi^*_{\mu} (z) \psi_{\nu} (z)
\end{equation}

\section{Derivation of $\Sigma^s$}
 $\Sigma^s$ is the IFR scattering self-energy treated in the self-consistent Born approximation (SCBA),
\begin{equation}\label{SelfEirs}
   \Sigma^{s}_{\mu{\bm k},\nu{\bm k''\!}}(\epsilon)
   =
  \langle V^{ifr}_{\mu\alpha} ({\bm k},{\bm k}_1) V^{ifr}_{\beta\nu} ({\bm k}_2,{\bm k''})\rangle
     G_{\alpha{\bm k}_1,\beta{\bm k}_2\!}(\epsilon)
\end{equation}
where $\epsilon$ is energy and $G$ is the full retarded Green's function. Repeated indices are summed. Plugging the matrix element $V^{ifr}$ in (\ref{VirpWS}) into (\ref{SelfEirs}), the coefficient in front of the Green's function becomes
\begin{equation}\label{VV}
\begin{aligned}
& \langle V^{ifr}_{\mu\alpha} ({\bm k} ,{\bm k}_1) V^{ifr}_{\beta\nu} ({\bm k}_2, {\bm k''}) \rangle \\
  = &\sum_{j}\frac{\delta E_j^2}{A^2} \iint\!\!d^2 {\bm r}_1  d^2 {\bm r}_2 e^{-i(({\bm k} - {\bm k}_1) {\bm r}_1+ ({\bm k}_2 - {\bm k''}) {\bm r}_2)}  \\
   &\cdot \iint d z  dz' \mathcal{F}_{\mu\alpha\beta\nu}(z,z')\\
   &\cdot \langle (\theta(\tilde{z}_j-\xi_j({\bm r}_1)) - \theta(\tilde{z}_j))  (\theta(\tilde{z}'_{j}-\xi_{j}({\bm r}_2)) - \theta(\tilde{z}'_{j}))  \rangle\\
\end{aligned}
\end{equation}
where $\mathcal{F}_{\mu\alpha\beta\nu}(z,z')=\psi_\mu^*(z) \psi_\alpha(z) \psi_\beta^*(z') \psi_\nu(z')$.
The expression of the configurational averaged term depends on the signs of $\tilde{z}_j$ and $\tilde{z}'_j$. For $\tilde{z}_j<0$ and $\tilde{z}'_j<0$, we have
\begin{equation}
\begin{aligned}
(\theta &(\tilde{z}_j-\xi_j({\bm r}_1)) - \theta(\tilde{z}_j)) (\theta(\tilde{z}'_{j}-\xi_{j}({\bm r}_2)) - \theta(\tilde{z}'_{j}))\\
    =& \left\{ \begin{array}{l}
    1, \quad\xi_j({\bm r}_1)<\tilde{z}_j {\rm\ and\ } \xi_j({\bm r}_2)<\tilde{z}'_j \\
    0, \quad {\rm otherwise}
    \end{array} \right.
\end{aligned}
\end{equation}
thus
\begin{equation}
\begin{aligned}
&\langle (\theta(-\xi_j({\bm r}_1)) - \theta(\tilde{z}_j))  (\theta(\tilde{z}'_{j}-\xi_{j}({\bm r}_2)) - \theta(\tilde{z}'_{j}))  \rangle\\
=& \int^{\tilde{z}_j}_{-\infty} d\zeta \int^{\tilde{z}'_j}_{-\infty}  d\zeta' f^{(2)}_{\xi, r}(\zeta,\zeta')
\end{aligned}
\end{equation}
where $f^{(2)}_{\xi, r}(\zeta,\zeta')$ is the joint probability density, which describes the probability density of $\xi_j({\bm r}_1)=\zeta$ and $\xi_j({\bm r}_2)=\zeta'$ where $r=|{\bm r}_1-{\bm r}_2|$.
Similar expression can be obtained for the other signs of $\tilde{z}_j$ and $\tilde{z}'_j$. Thus configurational averaging term can be expressed as
\begin{equation}\label{cor}
 \begin{aligned}
  & \langle\left(\theta(\tilde{z}_j-\xi_j({\bm r}_1)) - \theta(\tilde{z}_j)\right)\left(\theta(\tilde{z}'_j-\xi_{j}({\bm r}_2)) - \theta(\tilde{z}'_j)\right)\rangle \\
  &= sgn\left(\tilde{z}_j \tilde{z}'_j\right)\iint_{(\zeta,\zeta')\in D} d\zeta d\zeta' f^{(2)}_{\xi, r}(\zeta,\zeta')\\
  \end{aligned}
\end{equation}
with the domain of integration $D$:
\begin{equation}\label{D}
  \begin{aligned}
  &D= \{  \begin{array}{ll}
         (-\infty,\tilde{z}_j), & \tilde{z}_j<0 \\
         (\tilde{z}_j,\infty), & \tilde{z}_j>0 \\
         \end{array} \times \{ \begin{array}{ll}
         (-\infty,\tilde{z}'_j), & \tilde{z}'_j<0 \\
         (\tilde{z}'_j,\infty), & \tilde{z}'_j>0\\ \end{array}
  \end{aligned}
\end{equation}
We notice (\ref{cor}) only depends on ${\bm r}_1-{\bm r}_2$. Thus one of the ${\bm r}$ integrals in (\ref{VV}) leads to $\delta({\bm k}-{\bm k}_1+{\bm k}_2-{\bm k''})$, which cancels one of the momentum sum in (\ref{SelfEirs}). Now we define ${\bm p} = {\bm k}-{\bm k}_1={\bm p} = {\bm k}_2-{\bm k''}$, and we have,
\begin{equation}\label{VV1}
\begin{aligned}
    & \langle V^{ifr}_{\mu\alpha} ({\bm p}) V^{ifr}_{\beta\nu} (-{\bm p})\rangle \\
    = & \sum_{j} \frac{\delta E_j^2}{A} \int d^2 {\bm r} e^{-i {\bm p} {\bm r}} \iint\!d z dz' sgn(zz') \\
   &\cdot\iint_{(\zeta,\zeta')\in D}\!d\zeta d\zeta' f^{(2)}_{\xi, r}(\zeta,\zeta')  \mathcal{F}_{\mu\alpha\beta\nu}(z,z')\\
\end{aligned}
\end{equation}
Summing over ${\bm p}$ in the self-energy (\ref{SelfEirs}), we obtain the final expression of $\Sigma^s$,
\begin{equation}\label{SelfEirs1}
\begin{aligned}
   &\Sigma^{s}_{\mu{\bm k}\nu{\bm k''}}(\epsilon)\\
   =& \sum_{{\bm p}}\langle V^{ifr}_{\mu,\alpha} ({\bm p}) V^{ifr}_{\beta,\nu} (-{\bm p})\rangle    G_{\alpha,{\bm k}-{\bm p},\beta,{\bm k}''+{\bm p}}(\epsilon) \\
=& \int\!d^2 {\bm p} \sum_{j} \frac{\delta E_j^2}{4\pi^2} \int\!d^2 {\bm r} e^{-i {\bm p} {\bm r}} \!\iint\!d z d z'  \\
   & \cdot sgn(zz')\iint_{(\zeta,\zeta')\in D} d\zeta d\zeta' f^{(2)}_{\xi, r}(\zeta,\zeta')  \\
   &\cdot \mathcal{F}_{\mu\alpha\beta\nu}(z,z') \cdot G_{\alpha,{\bm k}-{\bm p},\beta,{\bm k}''+{\bm p}}(\epsilon)
  \end{aligned}
\end{equation}

It is noted that when IFR is sufficiently small, one can assume constant wave functions across the interfaces, i.e.\ $\psi(z)\approx\psi(z_j)$.
With this approximation, the $z$ integral in (\ref{VV}) can be performed inside the configurational averaging. Then the expression (\ref{VV}) immediately reduces to the traditional form, which contains the assumption of a purely \mbox{2-D} IFR potential \cite{Wacker2002PhysRevB.66.245314,Tsui_APL2007transport},
\begin{equation}\label{VVtrad}
\begin{aligned}
    & \langle V^{ifr}_{\mu,\alpha} ({\bm k},{\bm k}_1) V^{ifr}_{\beta,\nu} ({\bm k}_2,{\bm k}'')\rangle  \\
    & = \!\delta_{{\bm k}-{\bm k}_{1\!}+{\bm k}_2-{\bm k}''\!}\!\sum_{j}\!\frac{\delta E_j^2}{A} \! \int \!d^2 {\bm r} e^{-i ({\bm k}-{\bm k}_1\!) {\bm r}\!}
     \langle \xi_j(0) \xi_{j}({\bm r})\rangle  F_{\mu\alpha\beta\nu}
\end{aligned}
\end{equation}
where $F_{\mu\alpha\beta\nu} = \psi_\mu^*(z_j) \psi_\alpha(z_j) \psi_\beta^*(z_{j}) \psi_\nu(z_{j})$, and the expression for $\langle \xi_j({\bm r}_1) \xi_{j}({\bm r}_2)\rangle $ is found in (\ref{xi}).

%
